# NEURAL VOICE CLONING WITH A FEW LOW-QUALITY SAMPLES

*Sunghee Jung, Hoirin Kim*

School of Electrical Engineering, KAIST, Daejeon, South Korea
{sh.ee, hoirkim}@kaist.ac.kr

**ABSTRACT**

In this paper, we explore the possibility of speech synthesis from low quality found data using only limited number of samples of target speaker. We try to extract only the speaker embedding from found data of target speaker unlike previous works which tries to train the entire text-to-speech system on found data. Also, the two speaker mimicking approaches which are adaptation and speaker-encoder-based are applied on newly released LibriTTS dataset and previously released VCTK corpus to examine the impact of speaker variety on clarity and target-speaker-similarity .

## 1. INTRODUCTION

Along with the leap of neural network, speech synthesis technology has been greatly improved. Single speaker text-to-speech has met the naturalness that is hard to distinguish between human speech[1][2][3][4][5]. However, speech synthesis still has remaining problem of requiring high-quality, lab-recorded training corpus compared to other speech-related research area such as speech recognition or speaker recognition. Especially the need to personalize a text-to-speech (TTS) is not satisfied yet due to afore-mentioned reason. It is expensive to obtain high-quality audio samples of desired target speaker to the amount that is large enough to train TTS.

There have been several studies for the goal of utilizing low-quality data on speech synthesis. Voice loop has experimented to extend their TTS to wild corpus from YouTube with auto-transcribed text[6]. Also, In the study of global style token, noisy condition has been considered as one of 'styles' and transferred into 'clean style' using their approach to represent desired style with combinations of global style tokens[7]. Hsu et. Al has studied using variational autoencoder to hierarchically control the high-level feature such as noise condition and low-level features such as speaking rate[8]. Also, there has been efforts to segment a low-quality pod cast corpus by detecting the breathing sound in order to improve the quality of wild corpus[9]. There has also been effort to use SEGAN[10] for speech enhancement and use the enhanced corpus to train TTS[11] in order make decision about whether to include synthesized speech into anti-spoofing corpus[12], [13]. In [11], they make use of approximately 3 hours of publicly available low-quality data of the president Obama to train TTS and voice conversion models. On the other hand, in this research, we investigate the case where only a few low-quality data (6 utterances) are used to synthesize the voice of the target speaker.

Also, there have been works on better-representing desired personal characteristics. The works on multi-speaker TTS include Deep Voice 2, Deep Voice 3 and Voice Loop[14][15][6]. A step further from multi-speaker TTS, there have also been efforts to mimic or clone desired style or voice given one or a few samples. Works on voice cloning can roughly be categorized by Two. First, there is an approach of adaptation in which multi-speaker TTS is adapted with samples of target speaker. Second, there is an approach of using speaker or style embedding generated by auxiliary neural network. In this research, we will focus more on the latter. That is because the low-quality data, which we are interested in, is likely to have poor transcriptions and speaker-encoder-based approach of speaker mimicking does not require any transcriptions. Also, in this approach, cloning samples of target speaker does not affect linguistic embedding, thus preserving the clarity of original multi-speaker TTS. This speaker-encoder-based approach can be further categorized by how the multi-speaker TTS and speaker encoder interact. It can be explained in view of the training loss of speaker encoder. There are roughly three ways of training speaker encoder for voice cloning purpose. The first is using generative loss. This method has its advantage in that the loss used for training matches the actual purpose of the network, i.e., the speaker is trained to the goal of giving the speaker embedding that will match the mel spectrogram originally given as training input. The second option is discriminative loss[16][17]. In this method, speaker encoder is transfer-learned from pre-trained speaker verification model which is trained on discriminative loss such as triplet loss. The advantage of this approach is that unlike generative method, one can make use of speaker verification DB which is richer in speaker variability and lower in SNR. DB with low SNR tend to tamper clarity when used for training TTS but this way, low quality DB is only used for learning speaker embedding. The approach of using generative loss can only make use of the data that are clean enough to train multi-speaker TTS and it restricts the speaker variability of the training set for the speaker encoder.

There is also another type of loss called cyclic loss which constrains that speaker embedding extracted from synthesized speech must be similar to the speaker embedding extracted from original speech. [18] makes use of all three types of loss combined by weighted sum.

The rest of this paper is structured as following. Section 2 explains the overall system we developed to use wild DB as speaker mimicking target. Here, we used self-attention to better aggregate given cloning samples of the target speaker in wild. Section 3 describes the DB used for training each module described in section 2. Section 4 evaluates quality and cleanliness of enhanced speech and similarity and clarity of mimicked speech in subjective measures. Last but not least, we reach our conclusion at section 5.

## 2. SYSTEM DESCRIPTION

### 2.1. Multi-speaker TTS

In this system, we use deepvoice3-like multi-speaker TTS which is extended version of Tacotron 2 for supporting multi-speaker feature[15][3]. It deploys layers of stacked 1D convolutional gated linear unit to replace recurrent neural network and improve the parallelism at training time[18][19]. Detailed implementation of multi-speaker TTS match that of deep voice 3 [14]. It is trained with the sum of following three losses. L1 loss of linear and mel-spectrogram and 'done' loss which is implemented as binary cross entropy loss to predict last frame.

### 2.2. Speaker encoder

Here we implemented two versions of speaker encoder using generative loss. The difference of two encoders is the way speaker embeddings are aggregated on temporal domain. Speaker embedding is originally generated frame-by-frame. However, we need to aggregate them temporally to make one representation per one sample utterance of target speaker and then aggregate again along multiple cloning samples to obtain one representation per speaker.

The first implementation of speaker encoder(t1) to be introduced here aggregates in temporal domain using average pooling, while using self-attention to aggregate across cloning samples. The second implementation of speaker encoder(t2) uses self-attention for temporal aggregation as well as cross-samples aggregation. Fig. 1 and Fig. 2 are for the speaker encoders, respectively. Equations for this aggregation steps are included in appendix B and appendix C.

### 2.3. Speech enhancement

In this study, we utilized speech enhancement generative adversarial network (SEGAN) for speech enhancement. The implementation of SEGAN is available online[1]. This model is composed of generator and discriminator. Generator generates clean speech given noisy speech so that discriminator cannot tell whether it is clean speech from training set or cleaned speech which is output of generator. Loss is backpropagated through generator when the discriminator succeeds to distinguish the two so that generator can be updated in the direction of generating undistinguishable samples.

### 2.4. Vocoder

In this work, we used Griffin-Lim[20] and WORLD[21] vocoder. Neither of them are neural vocoder, resulting lower speech quality when compared to state-of-art vocoders. Vocoders play significant role when deciding the naturalness of synthesized speech. Thus the overall quality of the synthesized speech is low in this study.

## 3. CORPUS

### 3.1. Multi-speaker TTS, speaker encoder

Two multi-speaker TTS models are trained. First, 102 speakers are used for training TTS model 1. Six speakers from the VCTK are left out for test purpose. This first TTS model used WORLD vocoder for generating wav out of spectrum. Thus, converter part of WORLD is implemented and trained in the deep voice 3 structure.

For the second model, 460 hours of clean LibriTTS training set was used for multi-speaker TTS and speaker encoder. There are originally 1151 speakers in the train-clean (higher than 20 dB) dataset but since we train speaker encoder that generates speaker embedding out of 6 audio samples, 66 speakers who had less or equal to 6 samples were eliminated and total of 1085 speakers are used for training. The speakers who had exactly 6 samples are used for test purpose later. Each utterance was trimmed by energy-based voice activity detection (VAD) with the threshold of 60 dB.

Texts were used for multi-speaker TTS and not for speaker encoder. Each words were looked up in CMUdict[2] for its pronunciation. To improve the end-to-end ability of reading out-of-vocabulary words, some of the words were randomly forced to remain as grapheme even though it was found in the dictionary with probability of 0.5 as in [14].

### 3.2. Speech enhancement

For training speech enhancement, we use two separate settings. First, we train SEGAN with noisy VCTK. This contains 5 types of noise of 4 SNRs 17.5 dB, 12.5 dB, 7.5 dB and 2.5 dB to create 20 different noisy conditions.

---

[1] https://github.com/ssarfjoo/improvedsegan

[2] http://www.speech.cs.cmu.edu/cgi-bin/cmudict

Secondly, we train speech enhancement network with noisy, reverberant, noisy and reverberant, device-recorded VCTKs. Noisy condition here also includes 20 different conditions. For both settings, the VCTK speakers are 28 speakers using England accent and gender is equally balanced and there are about 400 utterances for each speaker.

This control was made to figure out whether absence of noise or distortion is more important when enhancing wild speech for the purpose of extracting speaker embedding of cloning target. As expected, the SEGAN model where it is trained with noise DB has lower distortion and is less clean, while the one trained with 4 different types of distortion removes more noise and results in more distortion. The effects of this control on speaker mimicking is described in next section.

## 4. EXPERIMENTS AND RESULTS

Table 1. Subjective evaluation of synthesized speech

|  | Similarity | Clarity |
|---|---|---|
| VCTK baseline | 2.00 | 1.50 |
| VCTK_adapt | 2.17 | 1.50 |
| VCTK_t1 | 2.33 | 3.00 |
| VCTK_t2 | 2.50 | 3.00 |
| LibriTTS baseline | 2.50 | 3.33 |
| LibriTTS_adapt | 2.50 | 2.67 |
| LibriTTS_t1 | 1.83 | 2.67 |
| LibriTTS_t2 | 1.67 | 2.67 |

Table 2. Subjective evaluation of enhanced speech and the mimicked speech which utilizes the enhanced speech

|  | Quality | Cleanliness |
|---|---|---|
| Original baseline | 4.17 | 2.50 |
| g1 | 2.00 | 3.17 |
| g2 | 1.17 | 2.50 |
| VCTK_t2_gx | 1.67 | 3.67 |
| VCTK_t2_g1 | 2.00 | 3.00 |
| VCTK_t2_g2 | 2.33 | 4.00 |

Table 3. The time required for enrolling new speaker for adaptation-based and speaker-encoder-based speaker mimicking

|  | Time for enrolling new speaker |
|---|---|
| Adaptation-based | 15 min. |
| Speaker-encoder-based | 11 sec. |

In this section, we conduct experiments to answer three following questions. First, we try to find out whether using self-attention for temporal aggregation improves the cloning quality or not. Secondly, we want to find out how cleanliness or distortedness of enhanced speech affects extracting speaker embedding for voice cloning purpose. Answer to these questions can help deciding hyperparameters or design of training set by giving hint about which aspect to focus among trades off of all speaker enhancement system- cleanliness and low distortion of the contents. Third, we want to verify whether adaptation-based mimicking or encoder-based mimicking gives better result under the condition where there are only a few (6) samples of target speaker available. Six samples of each target speaker are approximately length of 30 seconds in total.

Hyper parameter settings are mostly as in [22]. Detailed hyperparameter settings are included in Appendix D, E and F. for VCTK and LibriTTS.

Subjective tests were conducted on 6 listeners. Three of them are English-native speakers and other three of them are not (Korean-native speakers). Each subject listened to all the test sentences for evaluation.

At table 1, 'VCTK baseline' and 'LibriTTS baseline' each represent the synthesized speech of the speaker who were used for the training of each multi-speaker TTS model. All other items other than VCTK baseline and LibriTTS baseline are the score for the synthesized speech which mimic the voice of target speakers who are not used during training of multi-speaker TTS. Only 6 utterances of target speaker are used for all the mimicking experiments. T1 and t2 each represent the speaker encoder with average-pooling for temporal aggregation and self-attention for temporal aggregation, respectively. G1 and g2 represent the speech enhancement which is trained with noisy DB only and noisy, reverb, noisy reverb and device-recorded DB.

To begin with, we can see that similarity and clarity for VCTK baseline is very low. This is because the limited variability of speakers and sentences of VCTK. LibriTTS baseline gives much higher similarity and clarity because it is about 10 times bigger than VCTK. It can be seen that to train multi-speaker TTS with fair clarity, about a thousand of training speakers are required. Also, we can see from table 1, for VCTK, adaptation-based speaker mimicking gave approximately the same speaker similarity and the clarity with the multi-speaker baseline while speaker-encoder-based approaches gave increased similarity and clarity. This can interpreted that 6 cloning samples are enough for speaker mimicking. However, the fact that speaker mimicking result of VCTK_t1 and VCTK_t2 gave higher similarity and clarity than the baseline needs further thorough examination since it is against intuition. On the other hand, for LibriTTS, adaptation-based speaker mimicking gave approximately the same similarity as the baseline while encoder-based speaker mimicking gave lower similarity. We assume this is due to the characteristics of the mimicking target speaker in LibriTTS. Target speaker of LibriTTS mimicking task is reading children's book in an exaggerated way, and 6 samples that are used for cloning vary drastically in style. We assume it would have been much trickier to extract consistent speaker characteristic out of those six sentences which vary drastically in speaking style. For VCTK mimicking, the target speaker is reading in rather monotonous style and we assume this consistency among cloning samples of VCTK must have benefitted the speaker-encoder based mimicking approach. For comparing t1 and t2 speaker encoder, the speaker encoder type did not affect clarity for speaker mimicking and the effect on similarity differed by the dataset.

Table 2 shows the effect of training dataset for speech enhancement on the quality and cleanliness of enhanced speech and mimicked speech which used the enhanced speech for generating speaker embedding by speaker encoder type 2. 'g1' represents that noisy VCTK and clean VCTK pair is used for training speaker enhancement. 'g2' represents that noisy, reverb, noisy reverb, device recorded VCTK are paired with the clean counterpart and used for the training of speech enhancement. 'gx' represents that no speech enhancement is used before extracting the speaker embedding with the speaker encoder. Original baseline is evaluated to be of high quality despite of the low cleanliness coming from the back ground shutter noise and poor recording condition of the original found cloning samples from Obama's public speech[3]. Enhanced result of g1 and g2 both have lower quality than the original baseline because the content of the original speech is distorted during the enhancement. After the enhancement, the cleanliness is improved for g1 while the cleanliness for g2 is remaining as the same. This cleanliness score of g2 results from the fact that even though the g2 enhancement suppressed a lot of noise from the original speech, the newly introduced distortion has been perceived as new type of noise to the subjects. As a result, one can see that g1 enhancement obtained better subjective score for both quality and cleanliness aspect when the enhanced audio itself is being evaluated. However, when those three types of audio samples are used for extracting speaker embedding using the speaker encoder type 2 and entered to the multi-speaker TTS as speaker embedding, the result changes drastically. When the speaker embedding is extracted from found low-quality speech without enhancement, the quality of the mimicked speech generated by the multi-speaker TTS is so low that it was hard to understand the content. The linguistic information in the speech is almost gone, while the synthesized speech was cleaner than the baseline because the TTS was trained on clean, high-quality VCTK speech. When the low-quality found speech is enhanced using g1 speech enhancement, the quality of the mimicked speech was improved enough so that the content was understandable. Last but not least, when the speech enhance model which strongly suppressed the noise to the degree of distorting the original content was used for the mimicking process, the quality was similar to the ones that uses clean cloning samples. This is due to the fact that encoder-based speaker mimicking utilizes pretrained multi-speaker model for deciding the linguistic embedding for desired text and speaker embedding only affects the voice characteristic. Thus, the loss of content which commonly occurs during the speech enhancement does not affect the clarity of the synthesized speech. Another aspect worth mentioning is that even though the table 2 only shows the speaker mimicking result using encoder type 2, we also experimented with encoder type 1 and it seemed the effect of encoder type seemed more significant when the cloning samples are from enhanced low-quality data than the clean cloning samples. That is because, when speech is enhanced, the quality of the enhanced speech becomes uneven along the time, even introducing silence intervals in the middle of the speech, making the role of self-attention on temporal aggregation more noticeable.

Table 3 shows the time required for enrolling a new speaker who are not used during training for the purpose of mimicking. Adaptation-based approach took 15 minutes on the average for the multi-speaker TTS to converge. It has been verified by listening to the synthesized speech using intermediate epochs of adaptation and comparing them that there has not been overfitting due to unnecessary additional epochs. On the other hand, speaker-encoder-based approach took only 11 seconds to enroll a new speaker. That is because the enrollment step for this approach is mere the inference step of speaker encoder and no additional training is required. It can be seen that 81 times more time is required for the enrollment of a new speaker for the adaptation-based speaker mimicking. Also, the adaptation-based speaker mimicking requires storing all the network parameters whenever additional speaker needs to be mimicked. On the other hand, speaker-encoder method only requires memory for storing the speaker embedding which is 256 dimensions in this study.

## 5. CONCLUSION

In this study, we attempted a new approach of mimicking a target speaker only using limited number of low-quality samples, unlike previous studies which use either large amount of target speaker samples[11] or high quality samples[22]. Here we found that the combination of speech enhancement and speaker encoder can generate mimicked speech which has comparable quality to synthesized speech of the speaker who are seen during the training of the multi-speaker baseline. Also, we observed that the use of self-attention for temporal aggregation had more significant effect on the speaker embedding when the speaker embedding is generated from enhanced speech due to the distorted frames in the enhanced speech even though this fact does not have solid proof due to lack of experiment. We also think that speaker-encoder-based approach is more appropriate for utilizing low-quality data because if adaptation-based approach is used, the encoder part of the multi-speaker TTS will be contaminated by the input of low quality data and the clarity will be tampered. Also, low-quality found data is likely to have wrong transcriptions or no transcriptions at all which is required for adaptation-based approach.

---

[3]https://americanrhetoric.com/speeches/barackobama/barackobamafinalpressconference.htm

For future work, we aim to conduct more thorough experiments to prove these points and also use neural vocoder to improve the overall quality of the synthesized speech.

Appendix A. Structure of speaker encoder

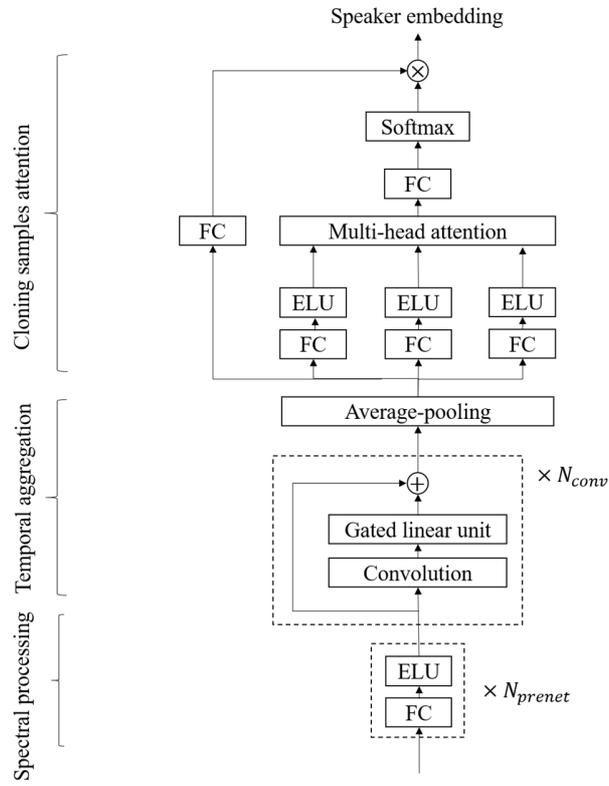

**Figure 1. Structure of speaker encoder 1.**

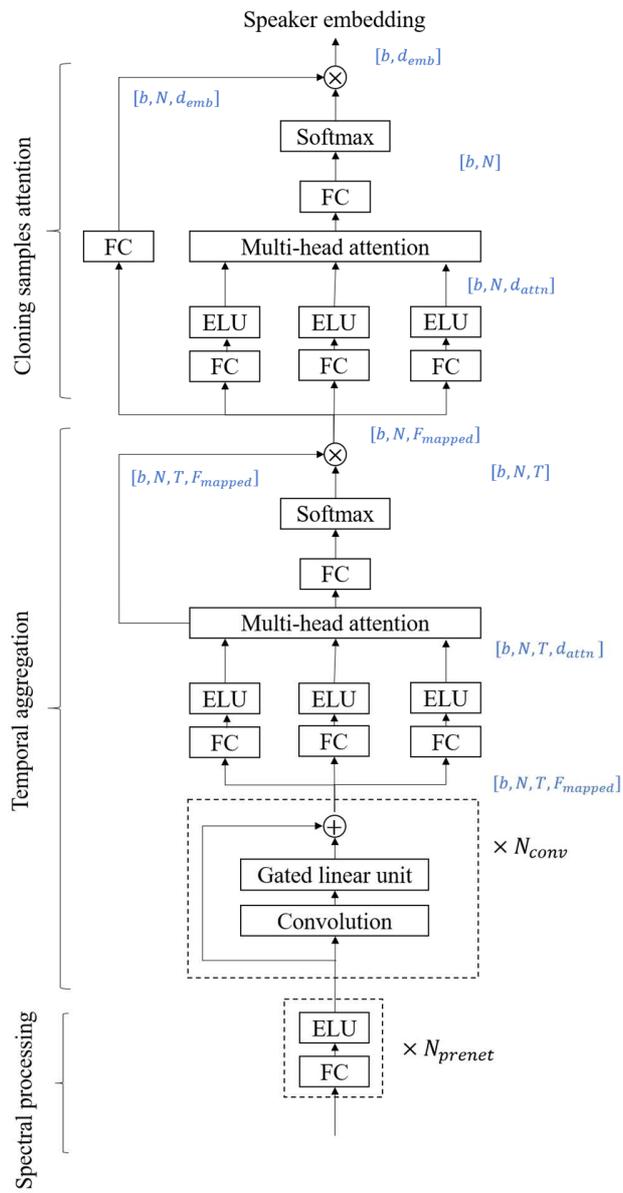

Appendix B. Equation for temporal self attention aggregation for speaker encoder.

For the every cloning sample, we need to obtain speaker embedding. Even though cloning sample index is unnecessary for explaining temporal aggregation, since it is conducted inside a single sample, but we include the cloning sample index for the continuity with the explanation of cross-sample aggregation given in next appendix.

The mel spectrogram has been transformed by spectral processing unit and the dimension has been changed from mel dimension $d_{mel}$ to feature dimension $f_{mapped}$. At temporal processing, we project this to $d_{attn}$ and apply exponential linear unit (ELU) for nonlinearity.

$$K'_j = ELU(y_j W'_{jK}) \in R^{T \times d_{attn}}$$
$$Q'_j = ELU(y_j W'_{jQ}) \in R^{T \times d_{attn}}$$
$$V'_j = ELU(y_j W'_{jV}) \in R^{T \times d_{attn}}$$

,where T is the number of frames.

Project above representation into smaller dimension $d_t = \frac{d_{attn}}{I}$ so that ensembling self-attention (i.e., multi-head attention) does not increase the number of parameters.

$$K_{ij} = y_j W_{ijK} \in R^{T \times d_t}$$
$$Q_{ij} = y_j W_{ijQ} \in R^{T \times d_t}$$
$$V_{ij} = y_j W_{ijV} \in R^{T \times d_t}$$
$$head_{ij} = softmax(\frac{Q_{ij} K_{ij}^T}{\sqrt{d_t}}) V_{ij}$$

, where $I$ is the number of attention heads and $i$ is index for head of multi-head attention. Multi-head can be considered as ensembles of context features. Heads are concatenated to form the original attention dimension $d_{attn}$, then projected to scalar, then normalized with softmax to return temporal self attention.

$$a_j = softmax(concat(head_0, \ldots, head_I) W_{jo}) \in R^t$$
$$e_j = \sum_{t=1}^{T} a_{tj} y_{tj}$$

, where product of $a_j$ and $y_j$ gives $e_j$, which is temporally aggregated speaker embedding for that cloning sample.

Appendix C. Equation for cloning sample attention aggregation for speaker encoder

Cloning sample attention part is exactly the same as *"Neural voice cloning with a few samples"*, however, since there is no detailed explanation, I would like to describe in detail here. To compare it with temporal aggregation above, the only difference is that here, cloning samples are projected to match speaker embedding dimension before being multiplied to the attention, whereas the feature of temporal attention was used as is. Therefore, the aggregated speaker embedding has desired $d_{embedding}$ dimension, which is the dimension of speaker embedding decided heuristically based on the variability of training dataset.

$$e_s: speaker\ embeddings\ from\ cloning\ samples \in R^{J \times f_{mapped}}$$

, where $J$ is the number of cloning samples and $f_{mapped}$ is the dimension of spectrogram after spectral processing unit which proceeded temporal aggregation.

$$K'_s = ELU(e_s W'_{sk}) \in R^{J \times d_{attn}}$$

$$K_s = K'_s W_{sk} \in R^{J \times d_t}$$

$$head_i = softmax(\frac{Q_{is} K_{is}^T}{\sqrt{d_t}}) V_{is}$$

$$a_j = softmax(concat(head_1, \ldots, head_I) W^o) \in R^J$$

$$e'_s = e_s w_s \in R^{J \times d_{embedding}}$$

, where $d_t = \frac{d_{attn}}{I}$ and $I$ is the number of heads. All the notations represent the same thing as in the Appendix B.

Appendix.D. Hyperparameters for training Multispeaker and speaker encoder for VCTK

```
{
 "name": "deepvoice3",
 "frontend": "en",
 "replace_pronunciation_prob": 0.5,
 "builder": "deepvoice3_multispeaker",
 "n_speakers": 108,
 "speaker_embed_dim": 256,
 "num_mels": 80,
 "fmin": 125,
 "fmax": 7600,
 "fft_size": 1024,
 "hop_size": 256,
 "sample_rate": 22050,
 "preemphasis": 0.97,
 "min_level_db": -100,
 "ref_level_db": 20,
 "rescaling": false,
 "rescaling_max": 0.999,
 "allow_clipping_in_normalization": true,
 "downsample_step": 4,
 "outputs_per_step": 1,
 "embedding_weight_std": 0.1,
 "speaker_embedding_weight_std": 0.05,
 "padding_idx": 0,
 "max_positions": 1024,
 "dropout": 0.050000000000000044,
 "kernel_size": 3,
 "text_embed_dim": 256,
 "encoder_channels": 512,
 "decoder_channels": 256,
 "converter_channels": 256,
 "query_position_rate": 2.0,
 "key_position_rate": 7.6,
 "key_projection": true,
 "value_projection": true,
 "use_memory_mask": true,
 "trainable_positional_encodings": false,
 "freeze_embedding": false,
 "use_decoder_state_for_postnet_input": true,
 "pin_memory": true,
 "num_workers": 12,
 "masked_loss_weight": 0.5,
 "priority_freq": 3000,
 "priority_freq_weight": 0.0,
 "binary_divergence_weight": 0.1,
 "use_guided_attention": true,
 "guided_attention_sigma": 0.4,
 "batch_size": 6,
 "adam_beta1": 0.5,
 "adam_beta2": 0.9,
 "adam_eps": 1e-06,
 "initial_learning_rate": 0.0005,
 "lr_schedule": "noam_learning_rate_decay",
 "lr_schedule_kwargs": {},
 "nepochs": 2000,
```

```
"weight_decay": 0.0,
"clip_thresh": 0.1,
"checkpoint_interval": 10000,
"eval_interval": 10000,
"save_optimizer_state": true,
"force_monotonic_attention": true,
"window_ahead": 3,
"window_backward": 1,
"power": 1.4,
"not_for_train_speaker": "300, 301, 302, 303, 304, 305",
"vocoder": "world",
"converter_dim": 187,
"cloning_sample_size": 6,
"f_mapped": 30,
"speaker_encoder_attention_num_heads": 8,
"speaker_encoder_attention_dim": 16,
"speaker_encoder_checkpoint_interval": 1000,
"vuv_weight_postnet": 0.4,
"spec_cmp_separator": "--"
}
```

Appendix E. Hyperparameters for training Multispeaker for LibriTTS

```
{
  "name": "deepvoice3",
  "frontend": "en",
  "replace_pronunciation_prob": 0.5,
  "builder": "deepvoice3_multispeaker",
  "n_speakers": 1151,
  "speaker_embed_dim": 256,
  "num_mels": 80,
  "fmin": 125,
  "fmax": 7600,
  "fft_size": 1024,
  "hop_size": 256,
  "sample_rate": 22050,
  "preemphasis": 0.97,
  "min_level_db": -100,
  "ref_level_db": 20,
  "rescaling": false,
  "rescaling_max": 0.999,
  "allow_clipping_in_normalization": true,
  "downsample_step": 4,
  "outputs_per_step": 1,
  "embedding_weight_std": 0.1,
  "speaker_embedding_weight_std": 0.05,
  "padding_idx": 0,
  "max_positions": 1024,
  "dropout": 0.050000000000000044,
  "kernel_size": 3,
  "text_embed_dim": 256,
  "encoder_channels": 512,
  "decoder_channels": 256,
  "converter_channels": 256,
  "query_position_rate": 2.0,
  "key_position_rate": 7.6,
  "key_projection": true,
  "value_projection": true,
  "use_memory_mask": true,
  "trainable_positional_encodings": false,
  "freeze_embedding": false,
  "use_decoder_state_for_postnet_input": true,
  "pin_memory": true,
  "num_workers": 12,
  "masked_loss_weight": 0.5,
  "priority_freq": 3000,
  "priority_freq_weight": 0.0,
  "binary_divergence_weight": 0.0,
  "use_guided_attention": true,
  "guided_attention_sigma": 0.4,
  "batch_size": 8,
  "adam_beta1": 0.5,
  "adam_beta2": 0.9,
  "adam_eps": 1e-06,
  "initial_learning_rate": 0.0005,
  "lr_schedule": "noam_learning_rate_decay",
  "lr_schedule_kwargs": {},
  "nepochs": 2000,
```

```json
"weight_decay": 0.0,
"clip_thresh": 0.1,
"checkpoint_interval": 10000,
"eval_interval": 10000,
"save_optimizer_state": true,
"force_monotonic_attention": true,
"window_ahead": 3,
"window_backward": 1,
"power": 1.4,
"not_for_train_speaker": "",
"vocoder": "",
"converter_dim": 513,
"cloning_sample_size": 6,
"f_mapped": 30,
"speaker_encoder_attention_num_heads": 8,
"speaker_encoder_attention_dim": 16,
"speaker_encoder_checkpoint_interval": 1000,
"vuv_weight_postnet": 0.4,
"spec_cmp_separator": "--",
"save_preprocessed_wav": "/home/admin/Music/preprocessed_libri_tts"
}
```

Appendix F. Hyperparameters for training speaker encoder for LibriTTS

```
{
 "name": "deepvoice3",
 "frontend": "en",
 "replace_pronunciation_prob": 0.5,
 "builder": "deepvoice3_multispeaker",
 "n_speakers": 1151,
 "speaker_embed_dim": 256,
 "num_mels": 80,
 "fmin": 125,
 "fmax": 7600,
 "fft_size": 1600,
 "hop_size": 400,
 "sample_rate": 22050,
 "preemphasis": 0.97,
 "min_level_db": -100,
 "ref_level_db": 20,
 "rescaling": false,
 "rescaling_max": 0.999,
 "allow_clipping_in_normalization": true,
 "downsample_step": 4,
 "outputs_per_step": 1,
 "embedding_weight_std": 0.1,
 "speaker_embedding_weight_std": 0.05,
 "padding_idx": 0,
 "max_positions": 1024,
 "dropout": 0.050000000000000044,
 "kernel_size": 3,
 "text_embed_dim": 256,
 "encoder_channels": 512,
 "decoder_channels": 256,
 "converter_channels": 256,
 "query_position_rate": 2.0,
 "key_position_rate": 7.6,
 "key_projection": true,
 "value_projection": true,
 "use_memory_mask": true,
 "trainable_positional_encodings": false,
 "freeze_embedding": false,
 "use_decoder_state_for_postnet_input": true,
 "pin_memory": true,
 "num_workers": 12,
 "masked_loss_weight": 0.5,
 "priority_freq": 3000,
 "priority_freq_weight": 0.0,
 "binary_divergence_weight": 0.0,
 "use_guided_attention": true,
 "guided_attention_sigma": 0.4,
 "batch_size": 2,
 "adam_beta1": 0.5,
 "adam_beta2": 0.9,
 "adam_eps": 1e-06,
 "initial_learning_rate": 0.0005,
 "lr_schedule": "noam_learning_rate_decay",
 "lr_schedule_kwargs": {},
 "nepochs": 2000,
```

```
  "weight_decay": 0.0,
  "clip_thresh": 0.1,
  "checkpoint_interval": 10000,
  "eval_interval": 10000,
  "save_optimizer_state": true,
  "force_monotonic_attention": true,
  "window_ahead": 3,
  "window_backward": 1,
  "power": 1.4,
  "not_for_train_speaker": "",
  "vocoder": "",
  "converter_dim": 513,
  "cloning_sample_size": 6,
  "f_mapped": 128,
  "speaker_encoder_attention_num_heads": 2,
  "speaker_encoder_attention_dim": 128,
  "speaker_encoder_checkpoint_interval": 1000,
  "vuv_weight_postnet": 0.4,
  "spec_cmp_separator": "--",
  "save_preprocessed_wav": "/home/admin/Music/preprocessed_libri_tts"
}
```